\documentclass[prl,amsmath,amssymb,floatfix,superscriptaddress,notitlepage,showpacs,twocolumn]{revtex4}
\usepackage{graphicx}
\usepackage{hyperref}
\usepackage{bm}
\usepackage{amssymb}

\begin{document}

\title{Entropic Measure for Localized Energy Configurations: 
Kinks, Bounces, and Bubbles}

\author{Marcelo Gleiser}
\email{mgleiser@dartmouth.edu}
\affiliation{Department of Physics and Astronomy, Dartmouth College,
Hanover, NH 03755, USA}

\author{Nikitas Stamatopoulos}
\email{nstamato@dartmouth.edu}
\affiliation{Department of Physics and Astronomy, Dartmouth College,
Hanover, NH 03755, USA}

\date{\today}

\begin{abstract}
We construct a configurational entropy measure in functional space. We apply it to several nonlinear scalar field models featuring solutions with spatially-localized energy, including solitons and bounces in one spatial dimension, and critical bubbles in three spatial dimensions, typical of first-order phase transitions. Such field models are of widespread interest in many areas of physics, from high energy and cosmology to condensed matter. Using a variational approach, we show that the higher the energy of a trial function that approximates the actual solution, the higher its relative configurational entropy, defined as the absolute difference between the configurational entropy of the actual solution and of the trial function. Furthermore, we show that when different trial functions have degenerate energies, the configurational entropy can be used to select the best fit to the actual solution. The configurational entropy relates the dynamical and informational content of physical models with localized energy configurations.

\end{abstract}
\maketitle

\section{Introduction}
Hamilton's principle of least action states that out of the infinitely many paths $x(t)$ connecting two fixed points in time, nature always chooses the one that leaves the action $S[x]$ stationary \cite{Hamilton}. This path, as is well known, is the solution to the Euler-Lagrange equation of motion, obtained from imposing that the action's first variation vanishes, $\delta S=0$. As we move from point particles to continuous systems, say, as described by a scalar field $\phi(x,t)$, Hamilton's principle will generate the partial differential equation (or equations, for vector- or tensor-valued fields) describing the motion that leaves the action functional $S[\phi]$ stationary. 
In this letter, we will explore the relation between information and dynamics, proposing an entropic measure that quantifies the informational content of physical solutions to the equations of motion and their approximations--the configurational entropy in functional space. As we will see, this measure may offer new insight into how spatially-localized ordered structures such as topological defects emerge in a widespread class of natural phenomena, from high-energy physics and cosmology \cite{Vilenkin} to condensed matter \cite{Gunton}.

To illustrate our point, we will investigate several nonlinear scalar field models in one and three spatial dimensions that have spatially-localized energy configurations such as solitons and critical bubbles. After defining the configurational entropy, we will compute it for models that admit spatially-localized energy solutions such as kinks, bounces, and critical bubbles. We will then compute the configurational entropy for several {\it ansatze} that approximate the solutions to the eom with varying degrees of coarseness. To compare solutions and approximations, we will define the relative configurational entropy--a measure of relative ordering in field configuration space--and show that it correlates well with the energy: the coarser the approximation to the solution, the higher its energy {\it and} its relative configurational entropy. 

We will also show how the relative configurational entropy can resolve ambivalent situations where the energies of different trial functions are degenerate, thus providing an efficient criterion to obtain optimized analytical approximations to the solutions of the equations of motion, as may be needed in different applications. On the basis of these results, we propose that nature is the ultimate optimizer, not only in extremizing energy through its myriad motions, but also from an informational perspective.

\section{Mathematical Preliminaries: Defining the Configurational Entropy}
Since we are interested in structures with spatially-localized energy, consider the set of square-integrable bounded functions $f(x) \in L^2({\bf R})$ and their Fourier transforms $F(k)$. Plancherel's theorem states that \cite{Fourier}
\begin{equation}
\int_{-\infty}^{\infty}|f(x)|^2dx = \int_{-\infty}^{\infty}|F(k)|^2 dk.
\end{equation}
Now define the modal fraction $f(k)$,
\begin{equation}
\label{modal fraction}
f(k) = \frac{|F(k)|^2}{\int|F(k)|^2 d^dk},
\end{equation}
where the integration is over all $k$ where $F(k)$ is defined and $d$ is the number of spatial dimensions. $f(k)$ measures the relative weight of a given mode $k$. For periodic functions where a Fourier series is defined, $f(k)\rightarrow f_n=|A_n|^2/\sum |A_n|^2$, where $A_n$ is the coefficient of the $n$-th Fourier mode.

We define the configurational entropy $S_C[f]$ as
\begin{equation}
\label{entropy for discrete modes}
S_C[f] = - \sum f_n \ln (f_n).
\end{equation}
In analogy with Shannon's information entropy, $S_S = - \sum p_i \log_2 p_i$, which represents an absolute limit on the best possible lossless compression of any communication \cite{Shannon}, we can think of the configurational entropy as providing the informational content of configurations compatible with the particular constraints of a given physical system. Note that when all $N$ modes $k$ carry the same weight, $f_n = 1/N$ and the discrete configurational entropy has a maximum at $S_C=\ln N$. If only one mode is present, $S_C = 0$. These limits motivate the definition of Eq. \ref{entropy for discrete modes}. Further motivation is given below.

For general, non-periodic functions in an interval $(a,b)$, we define the continuous configurational entropy $S_C[f]$
\begin{equation}
\label{configurational entropy}
S_C[f] = - \int {\tilde f}(k)\ln [{\tilde f}(k)] d^d k,
\end{equation}
where ${\tilde f}(k)=f(k)/f(k)_{\rm max}$ and $f(k)_{\rm max}$ is the maximum fraction, in most cases of interest given by the zero mode, $k=0$. This normalization guarantees that ${\tilde f}(k)\leq 1$ for all $k$.
We call the integrand ${\tilde f}(k)\ln [{\tilde f}(k)]$ the configurational entropy density. We note that $S_C[f]$ is similar to the Gibbs entropy of nonequilibrium thermodynamics, although the latter is defined for a statistical ensemble with microstates with probability $p_i=\exp(-E_i/k_BT)/\sum\exp(-E_i/k_BT)$, where $E_i$ is the energy of the $i$-th microstate and $k_B$ is Boltzmann's constant \cite{entropy}. Note also that the continuous configurational entropy can be made dimensionless by dividing it by $m^{d}$, where $m$ is the relevant mass or energy-scale of the system. (We set $c=\hbar=1$).

As an example, consider a Gaussian in $d$ dimensions, $f(r) = N\exp(-\alpha r^2)$, and its Fourier transform, $F(k) = \frac{N\exp(-k^2/4\alpha)}{[2\alpha]^{(d/2)}}$. Using the definition of Eq.~\ref{modal fraction},
\begin{equation}
\label{Gaussian modal fraction}
f(k) = \frac{\exp(-k^2/2\alpha)}{[2\pi\alpha]^{d/2}}.
\end{equation}
The zero mode carries the most weight. Eq.~\ref{configurational entropy} gives
\begin{equation}
\label{Gaussian entropy}
S_C(\alpha) = \frac{d}{2}(2\pi\alpha)^{d/2}.
\end{equation}
In the limit of a very spread-out Gaussian, $\alpha\rightarrow 0$, $S_C\rightarrow 0$, while for a highly-localized Gaussian $\alpha\rightarrow \infty$, $S_C\rightarrow \infty$. $S_C$ estimates the information required in $k$-space to build the function $f(x)$. In fact, defining the dispersion of a function $f(x)$ as 
\begin{equation}
\label{dispersion}
D_0(f) = \int_{-\infty}^{\infty}x^2|f(x)^2| dx,
\end{equation}
we know that for $f(x)$ and $F(k)$ normalized to unity, they satisfy the minimum uncertainty relation$~D_0(f)D_0(F)=1$.
We thus expect the configurational entropy of a Gaussian to represent an absolute minimum for spatially-localized functions $f(x)$ parameterized by the same spatial dispersion with one (or more) maxima for $x \in (-\infty,\infty)$. We can thus write, $S_C[f] \geq S_C(\alpha)$. 

In physical applications where the function $f(x)$ represents a spatially-confined structure (e.g. a $\phi^4$ or a sine-Gordon kink \cite{solitons}), the parameter $\alpha~$Ñor its equivalent determining the spatial extent of the structure~Ñwill be restricted by the relevant interactions or constraints. For example, a free particle of mass $m$ confined to a rigid-wall box of length $2R$ has momentum $p_n = n\hbar \pi/R$. If modeled by a Gaussian with dispersion $\sigma\equiv (2\alpha)^{-1/2}$ centered at $R$, the uncertainty relation gives $\alpha \leq \pi^2/2R^2$. On the other hand, $\sigma^2=(1/2\alpha)< R^2$ and so, $\alpha \geq 1/2R^2$. We thus have, $1/2 \leq \alpha R^2 \leq \pi^2/2.$ Since $2R$ cannot be smaller than the reduced Compton wavelength of the particle, $2R\geq 2\pi/m$, we get, $1/2\pi^2\leq \alpha/m^2 \leq 1/2$.

\section{Free Field in a Box}
As another application to motivate our definition of the configurational entropy, consider a free massive scalar field in a rigid-wall box of size $L$. Imposing $\phi(0,t)=\phi(L,t) =0$, we obtain the familiar quantization of momentum modes, $k_n = n\pi/L$, and dispersion relation $\omega^2_n = (n\pi c/L)^2 +m^2c^2$, with $n=1,2...$. Further imposing that $\dot\phi(x,0)=0$, the general solution is
$\phi(x,t) = \sum A_n\sin k_nx \cos\omega_n t$, where the coefficients $A_n$ are given in terms of the initial configuration $\phi_0(x) = \phi(x,0)$ as $A_n = (2/L)\int_0^L \phi_0(x)\sin [k_nx] dx$. Thus, the Fourier amplitudes $\{A_n\}$ determine the contributions from different normal modes to a given precision of the series expansion. The configurational entropy of a general field profile, obtained from Eq.~\ref{entropy for discrete modes}, is given by
\begin{equation}
\label{entropy normal modes}
S_C[\phi] = - \frac{1}{\sum |A_n|^2}\ln \left [\frac{\prod (|A_n|^2)^{(|A_n|^2)}}{(\sum |A_n|^2)^{(\sum |A_n|^2)}}\right ].
\end{equation}
If $A_n = A$ for $n=1,...,N$, then $S_C[\phi] = \ln N$, and the configuration entropy is maximal, as it should \cite{entropy}. Also, if only a single mode is present, $S_C[\phi] = 0$: the most ordered state is a pure state. From an initial configuration with a certain $S_C[\phi_0]$, the configurational entropy will evolve in time as the Fourier coefficients $A_n(t)=A_n\cos\omega_n t$ change.

\section{Configurational Entropy of Kinks}
Consider a scalar field model in $1d$ with energy density $\rho_V[\phi] = (\dot\phi)^2/2 +(\phi')^2/2 +V(\phi)$, where the dot and the prime denote time and spatial derivative, respectively. We are interested in situations where the Euler-Lagrange equations admit static solutions with localized energy density so that the configurational entropy is well-defined. That is, the energy density must be square-integrable even if the fields aren't. This is the case, for example, for the kink  solutions for the double-well and for the sine-Gordon potentials \cite{solitons}, which we now examine in turn.

\subsection{Case 1: Symmetric double-well potential}
If $V(\phi) = (\lambda/4)(\phi^2 - m^2/\lambda)^2$, the kink (or antikink) is the static solution interpolating the two minima of the potential at $\phi_{\rm min} = \pm(m/\sqrt{\lambda})$ as $x\rightarrow \pm \infty$, and is given by $\phi_k(x) = \pm(m/\sqrt{\lambda})\tanh(mx/\sqrt{2})$. The kink's energy density is $\rho[\phi_k](x) = (m^4/2\sqrt{\lambda}){\rm sech}^4(mx/\sqrt{2})$ and its energy is $E_k = \sqrt{8}/3m^3/\lambda\simeq 0.9428 m^3/\lambda.$ From the Fourier transform of the energy density, Eq.~\ref{modal fraction} gives the modal fraction
\begin{equation}
\label{kink modal fraction}
f(k) = \frac{35\pi k^2(4\alpha^2+k^2)^2}{2304\alpha^7} {\rm csch^2}\left(\frac{k\pi}{2\alpha}\right),
\end{equation}
where $\alpha\equiv m/\sqrt{2}$. Using ${\tilde f}(k)=f(k)/f(0)$ into Eq.~\ref{configurational entropy}, we obtain the configurational entropy for the $\phi^4$ kink,
$S_C[\phi_k] =  1.2167$. 
To check how this value compares to other monotonic configurations satisfying the same boundary conditions that approximate the kink we use a variational approach. As an illustration, consider the function $h(x) = (2/\pi)\tan^{-1}(\alpha x)$ expressed in units of $m/\sqrt{\lambda}$. The energy $E[h]$ is minimized for $\alpha_c = [24\zeta(3)/\pi^2]^{1/2}m\simeq 1.7097m$. With this value, the energy is $E[h] = 1.0884m^3/\lambda$ and the configurational entropy--computed, as for the kink, from the energy density--is $S_C[h] = 2.412 > S_C[\phi_k]$. Hence, for this first example, we see that the trial function $h(x)$ has both larger energy (as it should) and larger entropy than the kink solution.

Given that we cannot guarantee that any monotonic trial function will have lower CE than the solution to the eom (see, e.g., Case 2 next), we define the relative configurational entropy, $\Delta S_C[f,g]$, which compares the configurational entropy of trial functions to that of the solution of the eom
\begin{equation}
 \Delta S_C[f,g] = \frac{|S_C[f]-S_C[g]|}{S_C[g]},
\label{entropyDifference}
\end{equation}
where we take $S_C[f]$ to be the configurational entropy of the trial function and $S_C[g]$ that of the solution to the eom. This can be used together with a similar quantity for the energy to investigate if there is a direct correlation between the efficiency of a trial function in approximating the solution to the eom and its configurational entropy. 

In Table \ref{phi4 results} we show the results for $S_C$, $\Delta S_C[f,g]$, and $\Delta E[f,g]$ for the trial function $h(x)$ above and two other approximations to the kink solution. We also show the value of the extremized variational parameter $\alpha_c$ for each {\it ansatz}.
In all three cases, the coarser the ansatz--in the sense of having larger energy than the kink--the larger its configurational entropy and its relative configurational entropy. To confirm that $\Delta S_C$ gives a measure of fitness, we computed the fitting parameter $\chi^2 = \int [f(x) - \phi_k(x)]^2 dx$. As $\Delta S_C$ grows so does $\chi^2$.

\begin{table}[htdp]
\begin{center}
\begin{ruledtabular}
\begin{tabular}{c|c|c|c|c} 
{Ansatz}      &$\alpha_c[m]$&  {$\Delta E$}  &  ${\Delta S_C}$   & $S_C$\\ \hline
$\tanh(\alpha x)$ & $1/\sqrt{2}$ & 0 & 0 & $1.2167$\\ \hline
$  4\tan^{-1}[\exp(\alpha x)]/\pi-1$    &   1.1647   & $1.3\times10^{-3}$   &   $0.0756$  &$1.3086$ \\ \hline
\begin{tabular}[c]{@{}c@{}} $\exp[\alpha x]-1,x<0$\\$-\exp[-\alpha x]+1,x\geq 0$ \end{tabular}
       & 0.9574     & $1.55\times10^{-2}$   &   0.435     &$1.746$    \\ \hline
$  2\tan^{-1}[\alpha x]/\pi    $    & 1.7097     & 0.1447       &      0.9823          &$2.412$        
\end{tabular}
\end{ruledtabular}
\end{center}
\caption{Comparison of energy and configurational entropy of the $\phi^4$ kink and several trial functions.}
\label{phi4 results}
\end{table}%

\subsection{Case 2: sine-Gordon potential}
For another example, varying the $1d$ action with the sine-Gordon potential, $V_{\rm sG}(\phi) = \left [1-\cos(a\phi)\right ]/a^2$, gives the static kink (antikink) solution $\phi_{\rm sG}(x) = (4/a)\tan^{-1}\exp[\pm x/\sqrt{a}]$, with energy density $\rho_{\rm sG}[\phi] = 2V_{\rm sG}(\phi)$. Integrating over space we get $E_{\rm sG} = 8/a^2=8m^3/\lambda$.  We will set $a=1$ from now on. From the energy density, we obtain the modal fraction $f(k) = (3/8)\pi k^2{\rm csch}^2(k\pi/2)$, and Eq.~\ref{configurational entropy} gives the configurational entropy for the sine-Gordon kink, $S_C[\phi_{\rm sG}] = 1.1804$. 
We now proceed as before and compare these values to those for trial functions that approximate $\phi_{\rm sG}(x)$, computing the configurational entropies $S_C$ and $\Delta S_C$. Results are displayed in Table \ref{sineGordon results}:

\begin{table}[htdp]
\begin{center}
\begin{ruledtabular}
\begin{tabular}{c|c|c|c|c} 
{Ansatz}      &$\alpha_c[m]$&  {$\Delta E$}  &  ${\Delta S_C}$ & $S_C$   \\ \hline
$  4\tan^{-1}[\exp(\alpha x)]$    &   1.0   & $0$   &   $0$ & 1.1804       \\ \hline
$  \pi\tanh[\alpha x]+\pi$    &   0.6087   & $1.2\times10^{-3}$   &   $0.0667$ & 1.1017       \\ \hline
\begin{tabular}[c]{@{}c@{}} $\pi\exp[\alpha x],x<0$\\$-\pi\exp[-\alpha x]+2\pi,x\geq 0$ \end{tabular}      & 0.8173     & $8.3\times10^{-3}$   &   0.3181 &1.5559          \\ \hline
$  2\tan^{-1}[\alpha x]+\pi    $    & 1.4136     & 0.1024      &      0.7310 & 2.0432                      
\end{tabular}
\end{ruledtabular}
\end{center}
\caption{Comparison of energy and configurational entropy of the sine-Gordon kink and several trial functions.}
\label{sineGordon results}
\end{table}%

Again, we see that as the energy of the approximation to the solution to the eom increases, so does its relative configurational entropy. (Although $S_C$ for the ${\tanh}[\alpha x]$ ansatz is off by about $6.7\%$, $\Delta S_C$ correlates perfectly with $\Delta E$ and with $\chi^2$ in all cases.) Although we haven't proven that this correlation between lower energy and lower relative configurational entropy holds for all possible trial functions approximating the solutions, the examples analyzed here make for a compelling case: the configurational entropy offers an efficient measure of the ordering associated with localized energy configurations. Next, we extend our approach to bounces and critical bubbles associated with metastable decay in $1d$ and $3d$. We will see that not only the same correlation holds, but that the relative configurational entropy provides a way to break possible energy degeneracy between two or more trial functions.

\section{Configurational Entropy of Bounces and Bubbles}

\begin{figure}[htbp]
\includegraphics[width=\linewidth]{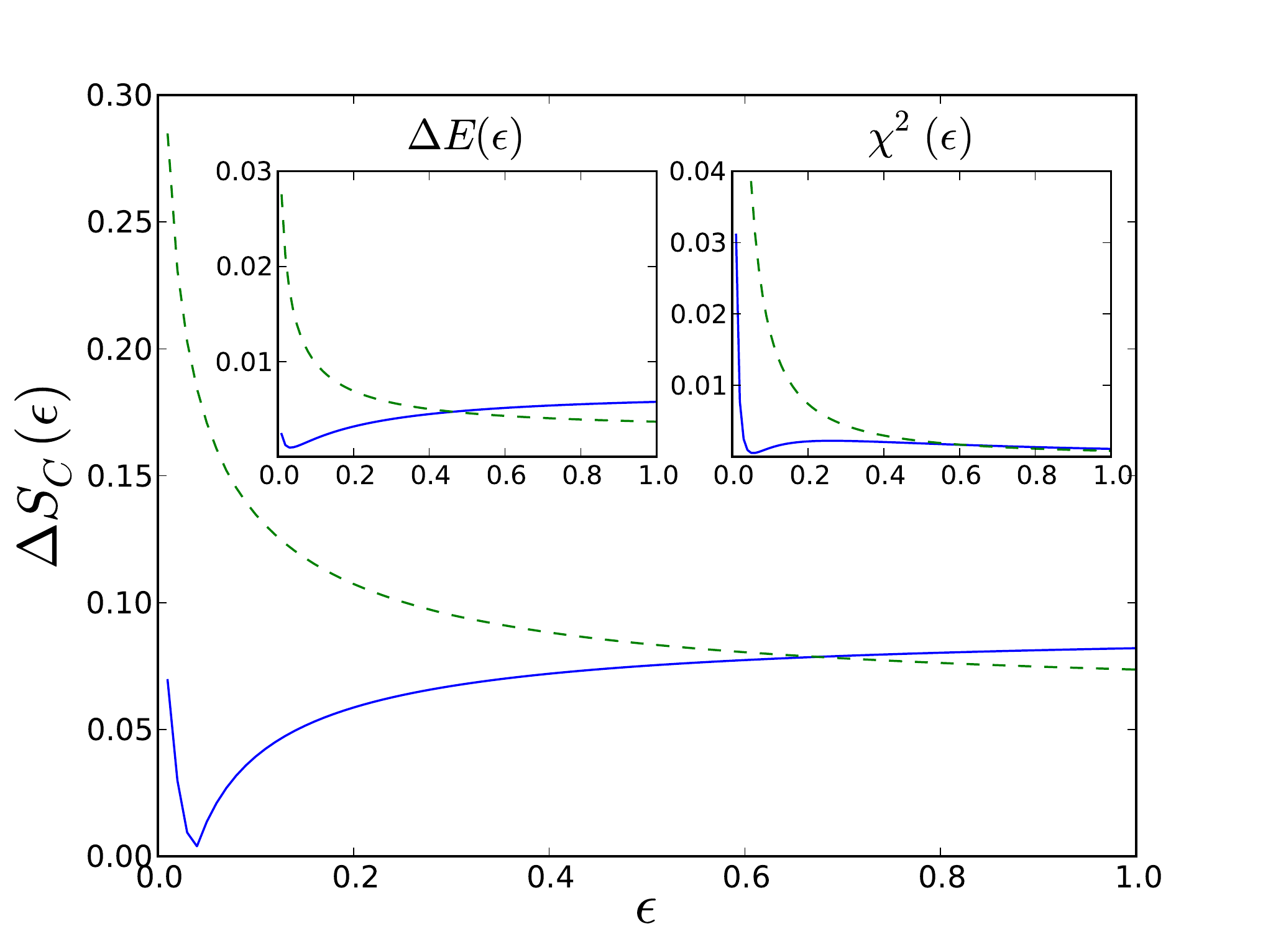}
\caption{Relative configurational entropy $\Delta S_R(\epsilon)$, energy difference $\Delta E(\epsilon)$ and $\chi^2(\epsilon)$  for two trial functions approximating the $1d$ bounce solution. Continuous line denotes the Gaussian and dashed line the sech. The energy difference is degenerate at $\epsilon_d=0.46$, while the Gaussian has lower relative configurational entropy for $\epsilon <0.68$.}
\label{1dbouncetrial}
\end{figure}

Consider a one-dimensional scalar field model with potential \cite{afg}
\begin{equation}
\label{bouncepot1}
V(\phi) = \frac{\lambda}{4}\phi^2\left (\phi - 2\phi_0\right )^2 -\alpha\phi_0\phi^3,
\end{equation}
where $\lambda$ and $\alpha$ are positive constants. For $\alpha=0$, the potential is a symmetric double-well as in the $\phi^4$ kink but shifted so that its minima are at $\phi=0$ and $\phi=2\phi_0$. Defining $\tilde{\phi} = \phi/\phi_0$, $\epsilon =\alpha/\lambda$ and dropping the tildes, the potential can be written as 
\begin{equation}
\label{bouncepot2}
V(\phi) = \frac{\lambda\phi_0^4}{4}\phi^2\left [\phi^2 -4\phi(1+\epsilon)+4\right ].
\end{equation}
The bounce [$\phi_{\rm bo}(x)$] is a static solution to the eom that vanishes as $x\rightarrow \pm\infty$ and reaches a maximum at $x=0$ at the classical turning point $\phi_{\rm tp}=2[(1+\epsilon)-(2\epsilon+\epsilon^2)^{1/2}]$ \cite{coleman}. So, differently from kinks, the field itself is square-integrable and it's simpler to compute its configurational entropy than that of the energy density, which has a double-peak profile. (One could still do the calculation using the energy density.) Also differently from kinks, solutions for different values of $\epsilon$ must be found numerically. The procedure, however, is the same: for each $\epsilon$, the bounce and its Fourier transform are found. Then, from Eq. \ref{configurational entropy}, the configurational entropy $S_C[\phi_{\rm bo}]$ is computed. To compare the bounce solution with trial functions that approximate it, we compute the relative configurational entropy $\Delta S_C[f,g]$ in two cases. A first trial function is the Gaussian $f(x)=\phi_{\rm tp}\exp[-\alpha(\epsilon) x^2]$, where $\alpha(\epsilon)$ is the variational parameter. Minimizing the energy with respect to $\alpha$ we find $\alpha_c(\epsilon) = 
2\sqrt{2}[\phi_{\rm tp}^2/8 - \phi_{\rm tp}(1+\epsilon)\sqrt{3}/3 +\sqrt{2}/2)]$. Note that for very small asymmetries (thin wall approximation) the Gaussian is not a good trial function for the bounce. In Fig. \ref{1dbouncetrial} we can see that this is nicely reflected in $\Delta S_C[f,g]$ for the Gaussian (continuous curve) which grows as $\epsilon\rightarrow 0$.

For a second trial function we use $f(x)=\phi_{\rm tp}{\rm sech}[\alpha(\epsilon)x]$. In Fig.~\ref{1dbouncetrial} we compare the energy (insert top left) and relative configurational entropy as a function of $\epsilon$ for the two trial functions. Notice that the energy is degenerate at $\epsilon_d=0.46$ and that for $\epsilon <[>] \epsilon_d$ the Gaussian [${\rm sech(x)}$] has smaller energy. However, the Gaussian has lower relative configurational entropy for all $\epsilon\leq 0.68$. At the degenerate point, the variational approach is ambiguous: energy considerations alone are insufficient to pick the best trial function there. The degeneracy is broken by using the relative configurational entropy, which clearly picks the Gaussian as the closest to the bounce solution at $\epsilon_d$. Should one trust the variational approach for $\epsilon>\epsilon_d$? According to the relative configurational entropy, only for $\epsilon\geq 0.68$. To resolve the impasse, we computed the best fit $\chi^2(\epsilon) \equiv\int [f(x)-\phi_{\rm bo}]^2dx$ for both trial functions as a function of $\epsilon$. The results are in the top right inset. For $\epsilon\leq 0.62$ (which includes $\epsilon_d$) the Gaussian is clearly a better approximation to the bounce. For $\epsilon>0.62$, the two trial functions have comparable values of $\chi^2(\epsilon)$. 

\begin{figure}[htbp]
\includegraphics[width=\linewidth]{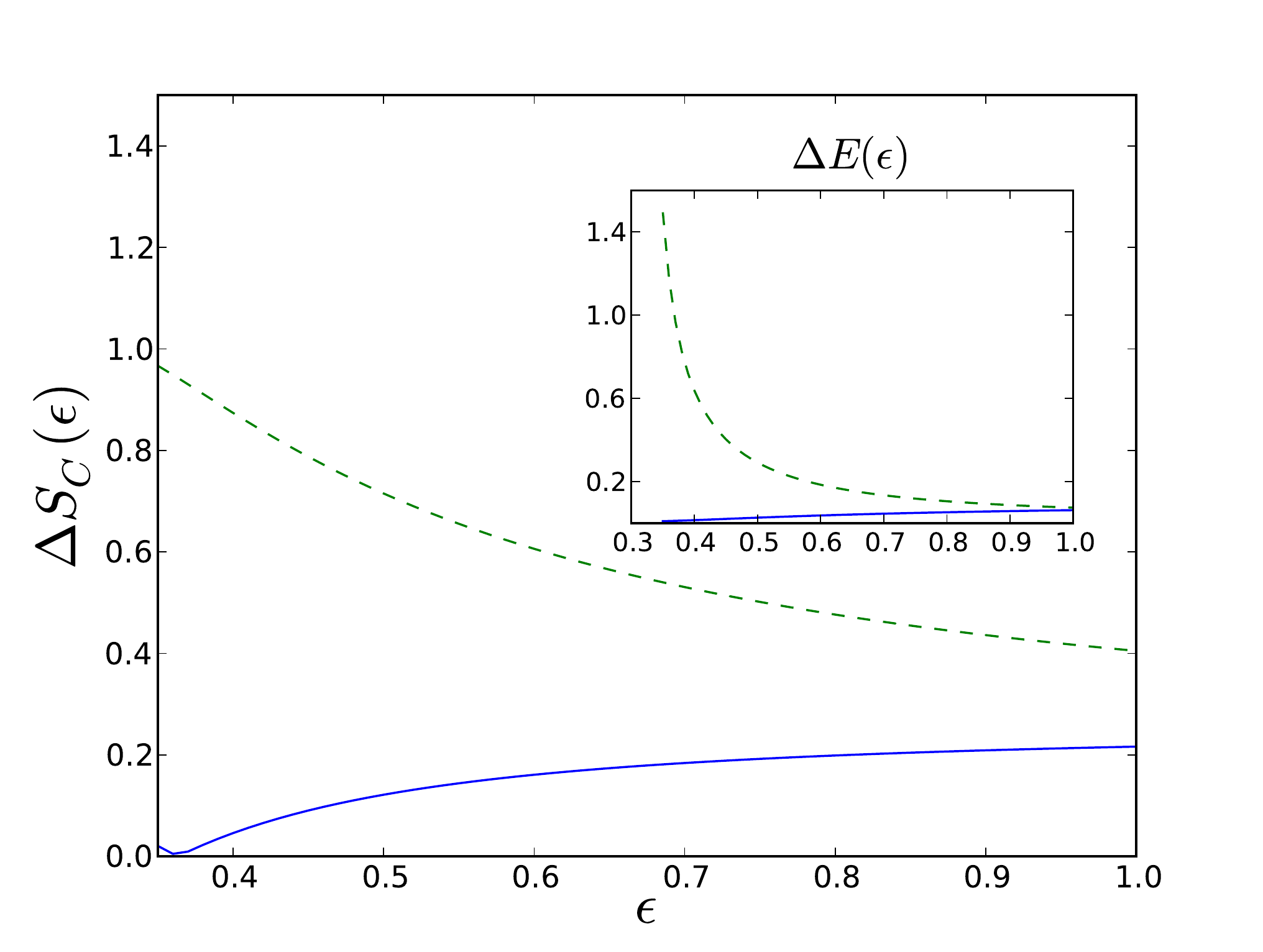} 
\caption{Relative configurational entropy $\Delta S_R(\epsilon)$ and energy difference $\Delta E(\epsilon)$ for two trial functions approximating the $3d$ critical bubble solution. Continuous line denotes the Gaussian and dashed line the sech. The Gaussian has lower relative configurational entropy and energy for all $\epsilon$ probed.}
\label{3dbubbletrial}
\end{figure}

We repeat the computations for a $3d$ scalar field model with the same potential as in Eq.~\ref{bouncepot1}. Now, the eom is $\phi'' +2\phi'/r = \partial V/\partial \phi$. The spherically-symmetric critical bubble solution has $\phi(r=0)=\phi_0$, $\phi'(r=0)=0$ and $\phi(r)=0$ as $r\rightarrow \infty$ \cite{falsevac}. As with the $1d$ case, we use the field to compute the relative configurational entropy as a function of the tilt $\epsilon$. The results are shown in Fig.~\ref{3dbubbletrial} for the same trial functions used in $1d$ and again show a correlation between energy minimization and ordering: the Gaussian has both lower energy and lower relative configurational entropy for all $\epsilon$ studied. For $\epsilon\lesssim 0.3$, we can't extremize the energy to find $\alpha_c$ since the critical bubble has a flatter profile near the origin. Other trial functions could be used to explore this range without difficulty.

\section{Summary and Outlook}
We propose an entropic measure of ordering in field configuration space for nonlinear models with spatially-localized energy solutions. We computed the relative configurational entropy $\Delta S_R$ for several trial functions approximating solutions to the eom for different models, showing that higher $\Delta S_R$ correlates with higher energy: solutions to the eom tend to be the most ordered, given their specific dynamic constraints. In cases where there is a degeneracy in the energy of trial functions, $\Delta S_R$ can be used as a discriminant. In forthcoming papers we will extend our approach to nonequilibrium field theory and cosmology. Our measure can be extended to models including gravity and gauge fields, and could potentially be used to discriminate between solutions in the superstring landscape \cite{landscape}.

MG is supported in part by a National Science Foundation grant PHY-1068027. NS is a Gordon F. Hull Fellow at Dartmouth College.

\end{document}